%% file: arxiv.tex
\newcommand{\CLASSINPUTtoptextmargin}{2cm}
\newcommand{\CLASSINPUTbottomtextmargin}{3.9cm}
\documentclass[conference,a4paper]{IEEEtran}
\usepackage{cite}
\usepackage{amsmath,amssymb,amsfonts}
\usepackage{graphicx}
\usepackage{textcomp}
\usepackage{xcolor}
\usepackage[oldcommands,boxed]{algorithm2e}

\makeatletter
\let\MYcaption\@makecaption
\makeatother

\usepackage[font=footnotesize]{subcaption}

\makeatletter
\let\@makecaption\MYcaption
\makeatother

\usepackage{enumitem}
\usepackage{stmaryrd}
\usepackage[utf8]{inputenc}
\def\BibTeX{{\rm B\kern-.05em{\sc i\kern-.025em b}\kern-.08em
    T\kern-.1667em\lower.7ex\hbox{E}\kern-.125emX}}
\DeclareMathAlphabet\EuScript{U}{eus}{m}{n}

\begin{document}

\title{Network Anomaly Detection based on\\Tensor Decomposition%*\\
}

\author{\IEEEauthorblockN{Ananda Streit, Gustavo  Santos, Rosa  Le\~{a}o,  Edmundo~de~Souza~e~Silva, Daniel  Menasché, Don Towsley\IEEEauthorrefmark{1}}
 \IEEEauthorblockA{Federal University of Rio de Janeiro, Rio de Janeiro, Brazil \quad \IEEEauthorrefmark{1}University of Massachusetts at Amherst, USA}
}

\maketitle

% http://www.peteryu.ca/tutorials/publishing/latex_captions

\captionsetup[subfigure]{labelformat=simple}
\renewcommand\thesubfigure{(\alph{subfigure})}

\input{tex/abstract.tex}

\begin{IEEEkeywords}
network measurement and analysis, machine Learning for networks, DDoS detection, tensor decomposition
\end{IEEEkeywords}

\input{tex/intro.tex}
\input{tex/related_work.tex}
\input{tex/tensor_decomposition.tex}
\input{tex/framework.tex}
\input{tex/residual.tex}
\input{tex/classification.tex}
\input{tex/clustering.tex}
\input{tex/conclusion.tex}

\textbf{Ackowledgments: } This work was partially supported by grants from CNPq, CAPES, FAPERJ and MCTIC/FAPESP and an NSF-MCTIC international cooperative grant.

\bibliographystyle{IEEEtran}
\bibliography{arxiv}
\end{document}

%% file: tex/abstract.tex
\begin{abstract}
The problem of detecting anomalies in time series from network measurements has been widely studied and
is a topic of fundamental importance.  Many anomaly detection methods are based on
packet inspection collected at the network core routers, with consequent
disadvantages in terms of computational cost and privacy.  We propose an alternative
method in which packet header inspection is not needed.  The method is
based on the extraction of a normal subspace obtained by the tensor
decomposition technique considering the correlation between different metrics.
We propose a new approach for online tensor decomposition where changes in the normal subspace
can be tracked efficiently.
Another advantage of our proposal is the interpretability of the obtained models.
The flexibility of the method is illustrated by applying it to two distinct
examples, both using actual data collected on residential routers. 
\end{abstract}

%% file: tex/intro.tex
\section{Introduction}
\label{sec:intro}

The problem of detecting anomalous events in computer networks has been widely studied
due to its relevance to network operation. However, these events are in general very hard to identify~\cite{chandola2009anomaly}.
The problem is challenging due to the wide variety of anomalies, low frequency of occurrences, and the definition of what is considered ``expected behavior".
An application example among the countless existing ones is detecting occasional changes in traffic patterns on a communication channel caused by a distributed denial of service (DDoS) attack.
DDoS attacks represent a major threat to proper network operation, wasting resources and creating network outages.
For instance, DDoS attacks targeted Amazon Web Services in October 2019 and were able to disrupt different services~\cite{amazonddos}.

In general, anomaly detection is based on the analysis of packet headers at the
core of the network, with potentially high computational cost and possible
privacy issues.  Our methodology differs from others in that it does not use
packet headers; it is based on distributed data collection at home routers, and
uses only a small amount of information.

The methodology is founded on \textit{tensor decomposition} to detect and diagnose
anomalous events using multivariate time series.
The approach was evaluated using time
series obtained from measurements collected at home routers of a medium-sized ISP.
Tensor decomposition allows the extraction of normal patterns from the metrics
considered, during different time intervals, and the identification of latent
relationships between them. We also devise a new online tensor decomposition
method that can efficiently track changes in the normal subspace.
The reported results show the effectiveness of the method to detect anomalies
in two different scenarios used as examples.
Nevertheless, we emphasize that the methodology is general and can be employed
in other scenarios.

This work shares similarities
with~\cite{lakhina2004diagnosing,lakhina2005mining},
where a normal subspace is defined by applying PCA and model residuals are used to detect
anomalies in a network.
In this work, extraction of the normal subspace is performed using the PARAFAC model~\cite{bro1997parafac},
which naturally allows the decomposition of multidimensional data and preserves relationships among
the metrics under evaluation.

The DDoS attack detection problem is the first example considered.
Based on observations of byte and packet upload/download counters which are
non-intrusive and requires no packet inspection,
we show that the proposed method can accurately 
detect attacks in both offline and online scenarios.
Using only measurements collected from home routers, the problem of identifying
time intervals within which performance degradation occurs constitutes the
second example. (Performance degradation is the anomaly in this case.)
The process can be easily
automated to identify and locate such anomalies and
analyze the quality of service in different parts of an ISP's topology.

\noindent {\bf Contributions}. 
Key contributions are summarized below: 

\noindent $\bullet$
\textit{Tensor decomposition to detect network anomalies.} 
Our framework is based on tensor decomposition.
We show that the PARAFAC model provides an interpretable
and efficient way to extract expected normal behavior,
taking into account the correlations among different metrics.

\noindent $\bullet$
\textit{New online tensor decomposition method.} 
Our method is based on a \emph{tensor window}~\cite{sun2008incremental}.
The results show the good accuracy and efficiency of the approach.

\noindent $\bullet$
\textit{Use of real data collected at home routers.}
We use time series obtained from
real network measurements collected at home routers to evaluate the framework.
Our method is capable of detecting different
types of anomalies based on simple metrics and without compromising users' privacy.

\noindent
\noindent $\bullet$
\textit{Use in different scenarios.}
The two application examples use different input metrics. 

Related work is presented in Section~\ref{sec:related_work}.
The tensor decomposition technique is discussed in Section~\ref{sec:tensor_decomposition}. 
Section~\ref{sec:framework} describes the proposed framework for anomaly detection and
we explain how the residuals are extracted in both offline and online
scenarios in Section~\ref{sec:residual}.
The DDoS attack detection example application is presented in Section~\ref{sec:classification},
and Section~\ref{sec:clustering} describes the second application example in which
network performance degradation intervals are detected.
Section~\ref{sec:concl} concludes our work.

%% file: tex/related_work.tex
\section{Related work}
\label{sec:related_work}

Anomaly detection methods are based on models that capture the normal behaviour
of the network~\cite{d2019survey}.
Most of the work in the literature of network anomaly detection is based on packet inspection at the core of the network~\cite{lakhina2004diagnosing, lakhina2005mining, maruhashi2011multiaspectforensics, xie2018graph, silveira2011astute},
which requires processing
privacy-sensitive information from packet headers, 
such as traffic volume between source and destination IPs and port number.
Recent work also employs packet inspection, but at home routers~\cite{doshi2018machine}.
Our work uses measurements performed by home routers without packet inspection,
providing a simple, efficient and privacy-preserving strategy.

Previous work by our group has also made use of measurements on home routers without packet inspection~\cite{gabriel_ddos}. 
Mendonça et al.~\cite{gabriel_ddos} focus on DDoS attacks that are detected with high accuracy
using only simple statistics from byte and packet counters obtained in a time window.
The current work uses another method (tensor decomposition)
and we show that it can be used to detect different types of anomalies. 
In addition, the results of our approach
are easier to interpret, since PARAFAC produces interpretable models~\cite{bro1997parafac}, and we are able to infer the normal daily behavior of a user, one of the main challenges in anomaly detection~\cite{chandola2009anomaly}.

Other previous works in the literature use subspace extraction methods (like PARAFAC) to detect network anomalies.
As an example, Maruhashi et al.~\cite{maruhashi2011multiaspectforensics} identify suspicious activities on the network
(such as port scanning and spreading of worms) by searching for abnormal subgraphs 
from the discovered patterns returned by a tensor model (PARAFAC).
Their method is heavily dependent on a manual choice of the patterns deemed interesting.
In addition, ~\cite{maruhashi2011multiaspectforensics} uses packet inspection,
considering a dataset with the format (source IP $\times$ destination IP $\times$ timestamp or port number).
This data structure is commonly used for network analysis with tensor models based on packet inspection.
In contrast, our work does not use data extracted from packet headers and consider tensors with the format
(user $\times$ network metrics $\times$ timestamp).

In the works of Lakhina et al.~\cite{lakhina2004diagnosing, lakhina2005mining},
the authors apply PCA to define the normal subspace.
Particularly, in~\cite{lakhina2005mining} anomalies that span multiple traffic features (metrics) are detected, 
similar to our work.
However, PCA is a matrix-based model and, unlike tensor-based models like PARAFAC,
it requires multidimensional data to be unfolded~\cite{sidiropoulos2017tensor} into a single, large matrix before its application.
The PARAFAC model, on the other hand, reveals the relationship between different metrics in multidimensional data,
making it more robust to noise. 
It also has the property of uniqueness, in contrast to PCA, where its inherent rotational freedom~\cite{bro1997parafac}
can lead to distinct interpretations concerning the structure of the normal subspace.

Xie et al.~\cite{xie2018graph} proposes an anomaly detection method using
a modified PARAFAC model that accounts for nonlinear data features.
The proposed algorithm considers the similarity of tensor slices in each mode
during the training process. In addition, residual anomalies are associated with a
sparse tensor and are isolated during the optimization process.
Kasai et al.~\cite{kasai2016network} also proposes a sparse tensor to account for abnormal flows.
However, both works ignore the interpretability of the model 
and evaluate the method using only artificially generated attack data taken
from arbitrary probability distributions, 
while we consider: (i) attack traffic generated by real malware and (ii) real performance degradation events.

We propose an online tensor decomposition approach based on the sliding window method of
Sun et al.~\cite{sun2008incremental}. Our solution has a lower computational cost while mantaining a good performance
for anomaly detection.
Kasai et al.~\cite{kasai2016network} also considers a tensor-based online algorithm.
Similar to our online algorithm, Kasai et al.~\cite{kasai2016network} uses the concept of sliding window 
and modifies the PARAFAC decomposition to deal with time and space complexities required for online approaches.
Other works also modify the PARAFAC decomposition for online application, but without the use of windows~\cite{nion2009adaptive,zhou2016accelerating}.
We focus on a simpler online solution, with a slight adaptation in PARAFAC decomposition.
In our approach the anomalies are detected by classifying or clustering the residuals obtained by tensor
decomposition.

%% file: tex/tensor_decomposition.tex
\section{Tensor decomposition}
\label{sec:tensor_decomposition}

In this section we briefly present a theoretical basis on tensor decomposition and describe our notation.
For details we refer to~\cite{sidiropoulos2017tensor}.
A tensor is a multidimensional matrix denoted by $\mathcal{X}$. 
We usually refer to the dimensions of $\mathcal{X}$ as \emph{modes}.
A third-order tensor $\mathcal{X} \in \mathbb{R}^{I \times J \times K}$ can be represented by a sum of three-way outer products~\cite{sidiropoulos2017tensor} as follows, 
\begin{align} 
  \mathcal{X} &= \mathcal{M} + \mathcal{E},    \quad \mathbf{a}_r \in \mathbb{R}^I, \mathbf{b}_r \in \mathbb{R}^J, \mathbf{c}_r \in \mathbb{R}^K  \\
     \mathcal{M}_{i,j,k} &= \sum^{R}_{r=1} \mathbf{a}_{r,i}  \mathbf{b}_{r,j} \mathbf{c}_{r,k}\textrm{,}     \label{eq:parafac} 
\end{align}
where $\mathcal{E}$ is the residual tensor and $R$ is the number of factors.
The \textit{factor matrices} (or loadings) define  model $\mathcal{M}$:
$A = $ $[\mathbf{a}_1, \mathbf{a}_2,  $ $\dots, \mathbf{a}_R] \in \mathbb{R}^{I \times R}\textrm{, }$ 
$B = $ $[\mathbf{b}_1, \mathbf{b}_2,  $ $\dots, \mathbf{b}_R] \in \mathbb{R}^{J \times R}\textrm{, }$  
$C = $ $[\mathbf{c}_1, \mathbf{c}_2,  $ $\dots, \mathbf{c}_R] \in \mathbb{R}^{K \times R}\textrm{.}$ 
Following standard notation, we also let $ \mathbf{a}_r =  A_{:,r}$, for $1 \leq r \leq R$, and
  $ \mathbf{a}^{(i)} =  A_{i,:}$, for $1 \leq i \leq I$.

The PARAFAC decomposition is obtained by minimizing the sum of squares of the residuals, 
i.e., the difference between $\mathcal{X}$ and $\mathcal{M}$.
Such difference is a nonconvex function; however, if we fix two of the factor matrices,
the problem is reduced to a linear least squares regression for the third matrix.
This is the basis of the \emph{Alternating Least Squares} (ALS) procedure~\cite{bro1997parafac}.
ALS estimates the factor matrices one at a time, keeping the others fixed.
The process iterates until a convergence criterion is satisfied or
there is no change in estimates.

In this work we use the method of \emph{Split-Half Validation} (SV)~\cite{harshman1984diagnostics}
in combination with \emph{Tucker Congruence Coefficient} (TCC)~\cite{lorenzo2006tucker} to estimate $R$ 
and evaluate whether the solution is unique and generalizable.

%% file: tex/framework.tex
\section{Framework}\label{sec:framework}

The proposed methodology consists of the following steps:

1) \textbf{Preprocessing:} 
	In the first step we perform data transformations needed to apply tensor decomposition, such as
    data scaling and filtering.
    
2) \textbf{Tensor Decomposition:}
	In this step we apply tensor decomposition to extract the normal subspace.
	We use the PARAFAC method due to the uniqueness of its solution and its capacity to deal with multivariate data~\cite{bro1997parafac}.

3) \textbf{Residual extraction:}
	The model obtained by tensor decomposition is used to extract the residuals and perform anomaly detection. 
	The idea is that anomalies are not well modeled by the normal subspace,
	allowing the separation between normal and anomalous behavior through residual analysis.

4) \textbf{Anomaly classification/clustering:}
	The final step varies depending on the application considered. 
	When the dataset contains labeled anomalies, we perform a supervised classification.
	On the other hand, there are applications where the labels for anomalies are unknown or hard to obtain.
	For these cases, we consider an unsupervised approach based on clustering.

%% file: tex/residual.tex
\section{Residual extraction}\label{sec:residual}

Our anomaly detection technique is based on analyzing the
PARAFAC residuals~\cite{bro1997parafac}.
Normal behavior is captured (modeled) by tensor decomposition
and anomalies are detected by investigating deviations from the modeled patterns.

\subsection{Offline residual extraction}

User data generally exhibits strong daily patterns over time. 
This leads us to split observations from users into independent daily series. 
We denote each of these series as a \emph{User-Day pair}, or \emph{UD pair}.

Let $I$ denote the number of UD pairs in our dataset.
We consider as inputs three-way tensors with modes UD pair~(factor matrix $A$),
metrics of interest~(factor matrix $B$) and time~(factor matrix $C$), 
denoted by indices $i$, $j$ and $k$, respectively.
Let $\mathcal{X}_{i,:,:}$ be  the $i$-th horizontal slice of tensor $\mathcal{X}$, i.e., 
 $\mathcal{X}_{i,:,:}$ is a two-dimensional matrix obtained by fixing the UD pair mode at value $i$~\cite{sidiropoulos2017tensor}.
Then, for each UD$_i$ with measurements  ${\mathcal{X}_{i,:,:} \in \mathbb{R}^{1 \times J \times K}}$,  
we obtain a model ${\mathcal{M}_{i,:,:} \in \mathbb{R}^{1 \times J \times K}}$ using PARAFAC ALS procedure. 
Residuals are measured as the difference between the model estimates 
and the input dataset ${\mathcal{E}_{i,:,:} =  \mathcal{X}_{i,:,:} - \mathcal{M}_{i,:,:}}$, 
where ${\mathcal{E}_{i,:,:} \in \mathbb{R}^{1 \times J \times K}}$.
As a UD refers to a day and our dataset consists of time series of one-minute bins, ${K = 1440}$. 

Next, we determine the residuals corresponding to measurements from new  UDs that were not previously
 used to parametrize  model~$\mathcal{M}$.
Let $\tilde{\mathcal{X}}_{\kappa,:,:}$ denote the measurements corresponding to a new UD$_\kappa$. 
 We use  factor matrices $B$ and $C$ from the previously trained model $\mathcal{M}$ (eq.~\eqref{eq:parafac}) 
  and the new measurements $\tilde{\mathcal{X}}_{\kappa,:,:}$  to
obtain  vector $\mathbf{\tilde{a}}^{(\kappa)} \in \mathbb{R}^{1 \times R}$.
Factor matrices $\tilde{A}$, B and C produce model $\mathcal{\tilde{M}}_{\kappa,:,:}$, with corresponding error 
 $\mathcal{\tilde{E}}_{\kappa,:,:}$, where $\tilde{A}_{\kappa,:} = \mathbf{\tilde{a}}^{(\kappa)}$.
Vector  $\mathbf{\tilde{a}}^{(\kappa)}$ is chosen to minimize quadratic error between model estimates and measurements. 
Let  $\tilde{\mathcal{X}}_{\kappa,:,:(1)}$  be the matrix unfolding
of tensor $\tilde{\mathcal{X}}_{\kappa,:,:}$ in its first mode~\cite{sidiropoulos2017tensor}, where $\tilde{\mathcal{X}}_{\kappa,:,:(1)} \in \mathbb{R}^{1 \times JK}$. Then,
\begin{align}
\mathcal{\tilde{M}}_{\kappa,:,:} = \mathcal{\tilde{\mathcal{X}}}_{\kappa,:,:} - \mathcal{\tilde{E}}_{\kappa,:,:} & {\Rightarrow}\mathbf{\tilde{a}}^{(\kappa)} (C \odot B)^T = \tilde{\mathcal{X}}_{\kappa,:,:(1)}  - \tilde{\mathcal{E}}_{\kappa,:,:(1)}  \nonumber  \\
&{\Rightarrow} \mathbf{\tilde{a}}^{(\kappa)} = \tilde{\mathcal{X}}_{\kappa,:,:(1)} ((C \odot B)^T)^\dagger \textrm{,}\label{eq:offline}
\end{align}
\noindent
where  $C \odot B$ denotes the Khatri-Rao product~\cite{sidiropoulos2017tensor}
between matrices $C$ and $B$  and 
$M^\dagger$ denotes the Moore-Penrose pseudo-inverse of matrix $M$~\cite{sidiropoulos2017tensor}.
Note that both $(C \odot B) \in \mathbb{R}^{JK \times R}$ and $((C \odot B)^T)^\dagger \in \mathbb{R}^{JK \times R}$.
As vector $\mathbf{\tilde{a}}^{(\kappa)}$ 
 minimizes the quadratic error,  
the corresponding  error ${\tilde{\mathcal{E}}}_{\kappa,:,:(1)}$  is orthogonal to  $((C \odot B)^T)^\dagger$ which implies~\eqref{eq:offline}.
Thus, the residuals of UD$_\kappa$ are obtained by
${\tilde{\mathcal{E}}_{\kappa,:,:} =  \tilde{\mathcal{X}}_{\kappa,:,:} - \tilde{\mathcal{M}}_{\kappa,:,:}}$, where 
 model ${\tilde{\mathcal{M}}_{\kappa,:,:} \in \mathbb{R}^{1 \times J \times K}}$ 
contains the new factor vector $\mathbf{\tilde{a}}^{(\kappa)}$.

Note that the  measurements corresponding to  UD pairs are  available by the end of a day, and 
residuals must be computed at that time.  
In addition, as  network conditions may change over time,
it is necessary to periodically  check if $\mathcal{M}$ is still a good 
model (e.g., using Split-Half validation~\cite{harshman1984diagnostics}). Otherwise,   $\mathcal{M}$ must be 
retrained to compute residuals for new UD pairs.

\subsection{Online residual extraction}

The online method tracks changes in the data 
by continuously recomputing the model using PARAFAC.
In the online scenario, time is divided into one minute slots and new data from all home routers is processed at every slot.
The online decomposition considers USERS instead of UD pairs as one of the tensor modes, 
and obtains a three-way tensor with modes user (mode $A)$, the metric of interest (mode $B$), and time (mode $C$). 
As soon as a new data stream arrives (every minute), the model is updated and residuals are extracted.

\begin{figure}[ht]
\centering
\subcaptionbox{FWO\label{fig:FWO}}{\includegraphics[width=43mm]{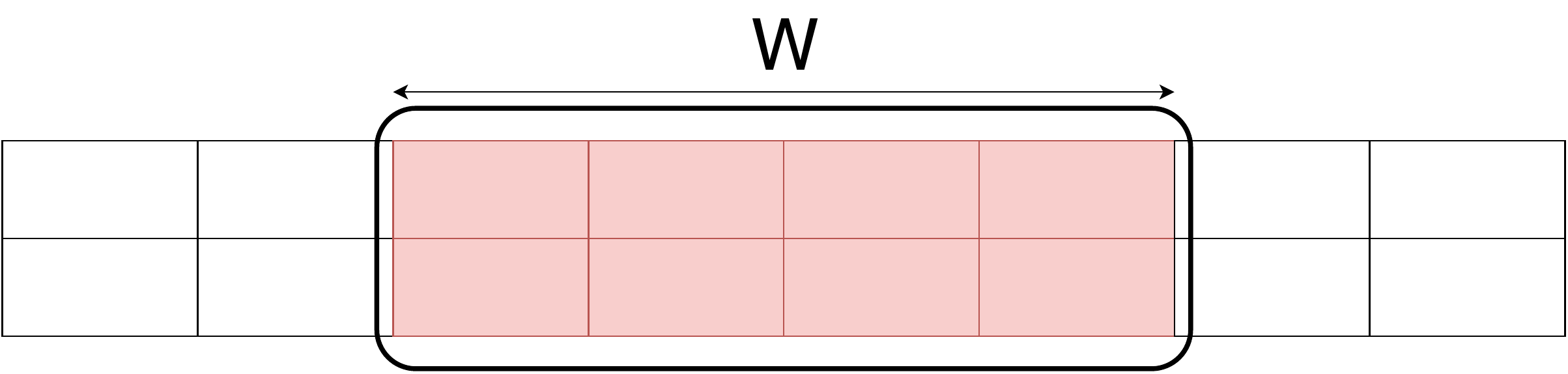}}
\subcaptionbox{PWO\label{fig:PWO}}{\includegraphics[width=43mm]{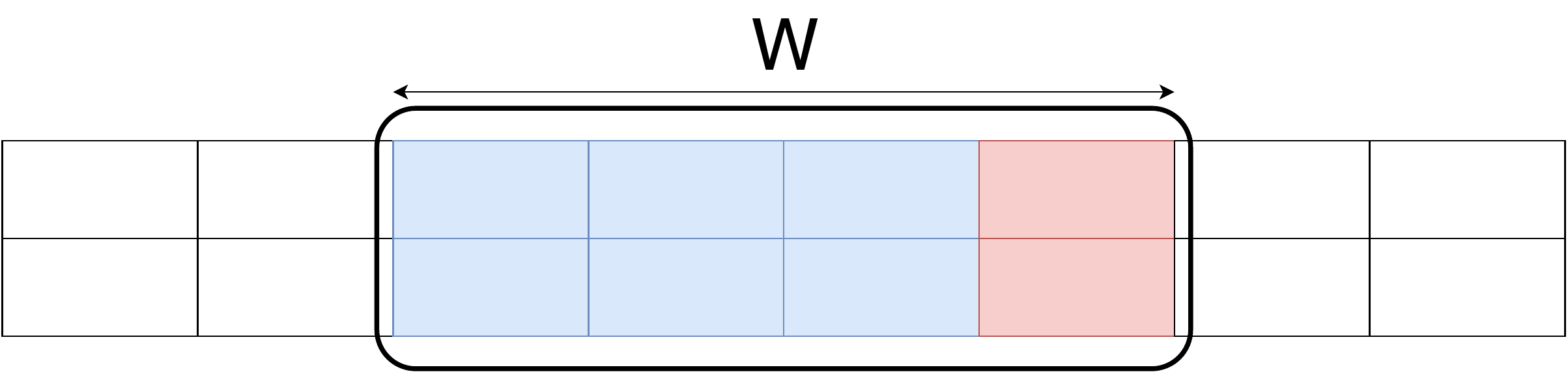}}
\caption{Online tensor decomposition approaches ($W = 4$). Red time slots are used to compute factor matrices
 $A$, $B$ and $C$. Blue time slots are used to compute  factor matrices $A$ and $B$.}\label{fig:online_tensor}
\end{figure}

We consider two different online residual extraction schemes. First, we describe 
Full Window Optimization (FWO)~\cite{sun2008incremental}.
Then, we propose Partial Window Optimization (PWO),
a simpler and more efficient FWO variant.
Figure~\ref{fig:online_tensor} 
illustrates the difference between the methods, as discussed below.
Note that both schemes allow the expansion of modes $A$ and $B$ throughout online decomposition, 
in case new users are added or new metrics of interest are collected, respectively.
In the offline model we denote by ${\mathcal{M}}$  (resp., $\tilde{\mathcal{M}}$) the model obtained before (resp., after) collecting additional measurements. In the online model variable $t$ already subsumes the number of collected samples, so we drop tilde from all variables.

\subsubsection{Full Window Optimization (FWO)~\cite{sun2008incremental}}

A simple approach to online tensor decomposition is based on a
\emph{tensor window}~\cite{sun2008incremental} ${\mathcal{X}(t, W) \in \mathbb{R}^{I \times J \times W}}$
over the time mode (mode $C$), where $W$ refers to the window size. 
At every minute $t$ the window slides and a new tensor is formed by combining
the $W-1$ previous slices $\{{\mathcal{X}_{:,:,t-W+1}, ..., \mathcal{X}_{:,:,t-1}}\}$
and the newly obtained data stream ${\mathcal{X}_{:,:,t} \in \mathbb{R}^{I \times J \times 1}}$,
representing a new lateral slice.
Since our network data presents strong daily patterns and we consider minute time slots,
we define the window size $W = 1440$.

To obtain models in FWO we use the same optimization method applied in the offline scenario for each sliding window.
Namely, for each window we compute a PARAFAC model ${\mathcal{M}(t, W) \in \mathbb{R}^{I \times J \times W}}$ 
using ${\mathcal{X}(t, W) \in \mathbb{R}^{I \times J \times W}}$ as input for the PARAFAC ALS algorithm.
Residuals are measured as the difference between the model 
and the input dataset ${\mathcal{E}(t, W) =  \mathcal{X}(t, W)-\mathcal{M}(t, W)}$. 
Usually we are interested in analyzing the behavior of the most recent sample.
Therefore, we consider the residuals of the last minute $t$,
${\mathcal{E}_{:,:,t} \in \mathbb{R}^{I \times J \times 1}}$.

A good initialization for the optimization algorithm can reduce the number of iterations needed to converge~\cite{sun2008incremental}.
Therefore, to speed up convergence, after we move the window forward, we initialize
the ALS algorithm with the previous model estimates, 
i.e., $\mathcal{M}(t-1, W)$ with factor matrices 
${A(t-1) \in \mathbb{R}^{I \times R}}$, ${B(t-1) \in \mathbb{R}^{J \times R}}$ and ${C(t-1) \in \mathbb{R}^{W \times R}}$.

FWO requires the computation of a whole new PARAFAC model at every window.
As such, it may not be suitable for online applications,
often requiring a high and variable number of iterations~\cite{nion2009adaptive}.
Our results indicate that this method is computationally expensive to be used  online for our application (see Figure~\ref{fig:time_compare}).
Hence, we propose a variation of this method to decrease computational cost while maintaining good performance.

\subsubsection{Partial Window Optimization (PWO)}

In order to reduce the run time we propose a modification to FWO
to obtain model $\mathcal{M}(t, W)$ as follows.
Consider the factor matrix related to the time mode
$C(t) \in \mathbb{R}^{W \times R}$
used to model the tensor window $\mathcal{X}(t, W)$.
To obtain $C(t)$ we keep the previous $W-1$ known loadings
$\{\mathbf{c}(t-W+1), ..., \mathbf{c}(t-1)\}$ fixed and
 compute the time mode loadings related to the last sample, i.e., we compute $\mathbf{c}(t)$.
The other factor matrices $A(t)$ and $B(t)$ are fully recomputed based on the tensor window
$\mathcal{X}(t, W)$.

Let $\mathcal{X}_{:,:,t}$ denote the measurements of a newly obtained data stream at time $t$.
The model is estimated by updating the unknown variables ($A(t)$, $B(t)$ and $\mathbf{c}(t)$, see Figure~\ref{fig:PWO}) 
alternately and iteratively,  
until a convergence criterion is satisfied or there is no change in estimates (Algorithm~\ref{algo:valit}).
As in the ALS algorithm,  
 matrices $A(t)$ and $B(t)$  and vector  $\mathbf{c}(t)$ 
are calculated by minimizing the quadratic error between model estimates and measurements. The sequence of updates is given by  lines 4-7 in Algorithm~\ref{algo:valit}.
\input{tex/italgorithm.tex}

In Algorithm~\ref{algo:valit}, $ \mathcal{X}_{(1)} $ and  $ \mathcal{X}_{(2)} $ are the tensor unfoldings of $ \mathcal{X}$ 
in its first and second modes, respectively,  $ \mathcal{X}_{(1)} \in \mathbb{R}^{I \times JK}$, $ \mathcal{X}_{(2)} \in \mathbb{R}^{J \times IK}$.  
Note that $\mathbf{c}(t) \in \mathbb{R}^{1 \times R} $, 
$A(t) \in \mathbb{R}^{I \times R}$ and $B(t) \in \mathbb{R}^{J \times R}$.  
 As with the FWO scheme, the factor matrices of model $\mathcal{M}(t, W)$ 
are initialized with the model estimates $\mathcal{M}(t-1, W)$ obtained for the previous window.
The residuals related to $t$
are obtained by ${\mathcal{E}(t, W)_{:,:,t} =  \mathcal{X}(t, W)_{:,:,t} - \mathcal{M}(t, W)_{:,:,t}}$.

%% file: tex/italgorithm.tex
\incmargin{1em} 
\restylealgo{boxed} \linesnumbered
\begin{algorithm}[b]
	 \begin{small}
		{
\SetLine
\SetLine
  $    A{(t)} \leftarrow A{(t-1)}, {\hskip0.5em\relax}
    B{(t)} \leftarrow B{(t-1)}$ \\
    $  C{(t)} \leftarrow  [\mathbf{c}{(t-W+1)}^T, \ldots, \mathbf{c}{(t-1)}^T, \mathbf{c}{(t-W)}^T]^T $ \\
  \While {not converged}{
 $\mathbf{c}{(t)} \leftarrow \mathcal{X}_{:,:,t(3)} ((B{(t)} \odot A{(t)})^T)^\dagger $ \\
 $C{(t)}_{W,:} \leftarrow \mathbf{c}{(t)}$ \\
$A{(t)} \leftarrow \mathcal{X}_{(1)} ((C{(t)} \odot B{(t)})^T)^\dagger$ \\
         $B{(t)} \leftarrow \mathcal{X}_{(2)} ((C{(t)} \odot A{(t)})^T)^\dagger$
   } 
{\bf return } $A(t), B(t), \mathbf{c}{(t)}$  
\caption{ Online algorithm}
\label{algo:valit}
\normalsize}
\end{small}
\end{algorithm}

%% file: tex/classification.tex
\begin{figure*}[tb]
\centering
\subcaptionbox{\label{fig:traffic_down}}{\includegraphics[width=45.3mm]{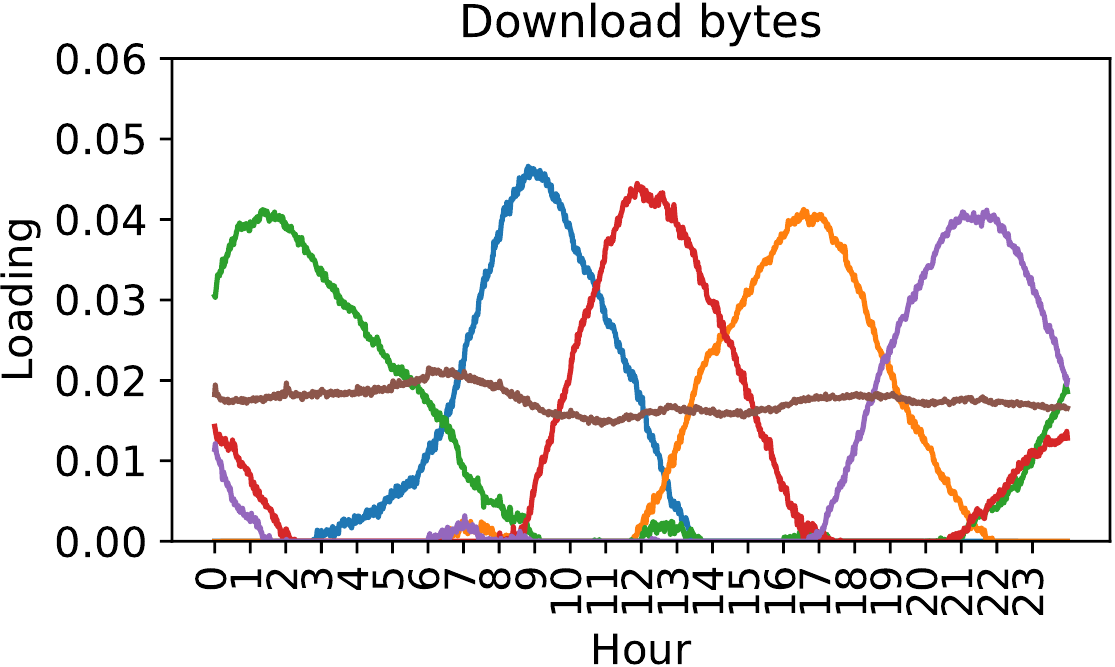}} 
\subcaptionbox{\label{fig:traffic_up}}{\includegraphics[width=44.2mm]{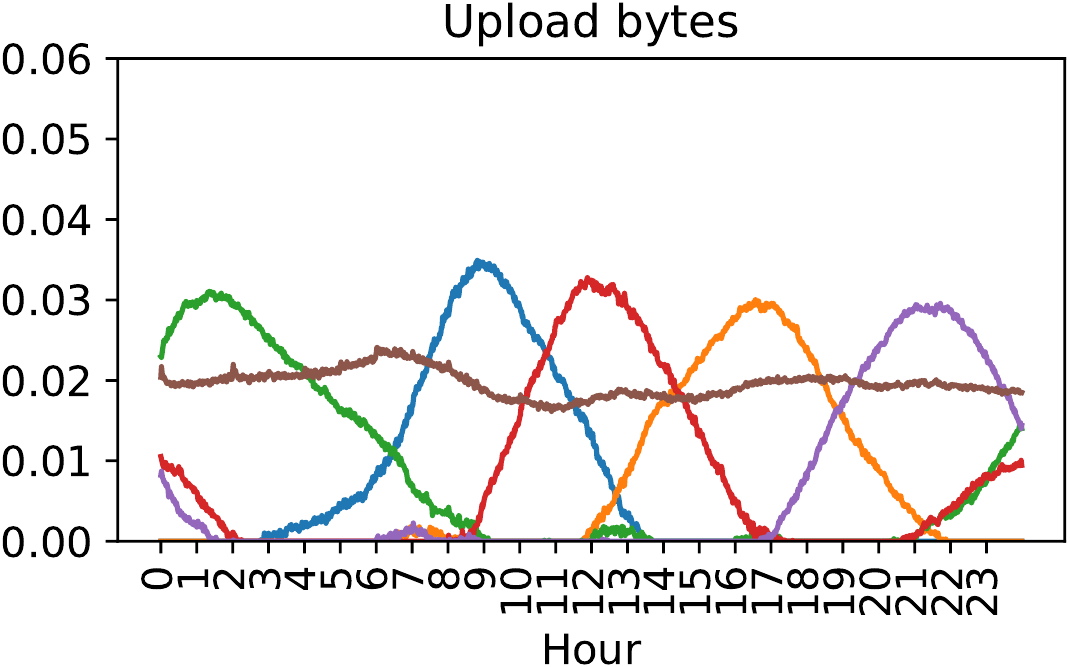}} 
\subcaptionbox{\label{fig:pkts_down}}{\includegraphics[width=44.2mm]{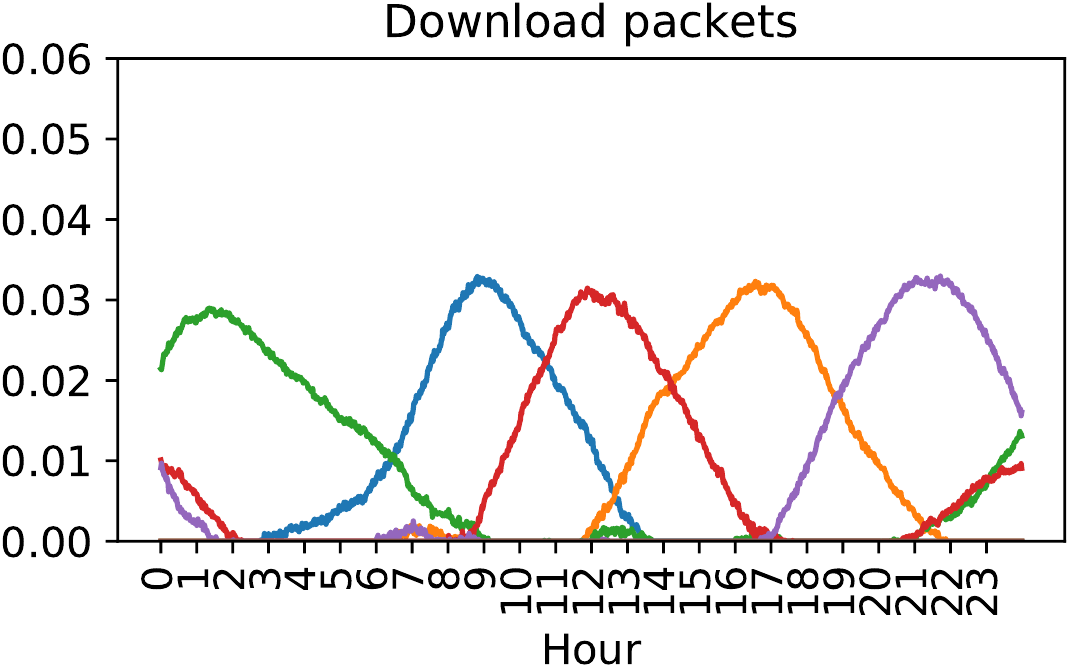}} 
\subcaptionbox{\label{fig:pkts_up}}{\includegraphics[width=44.2mm]{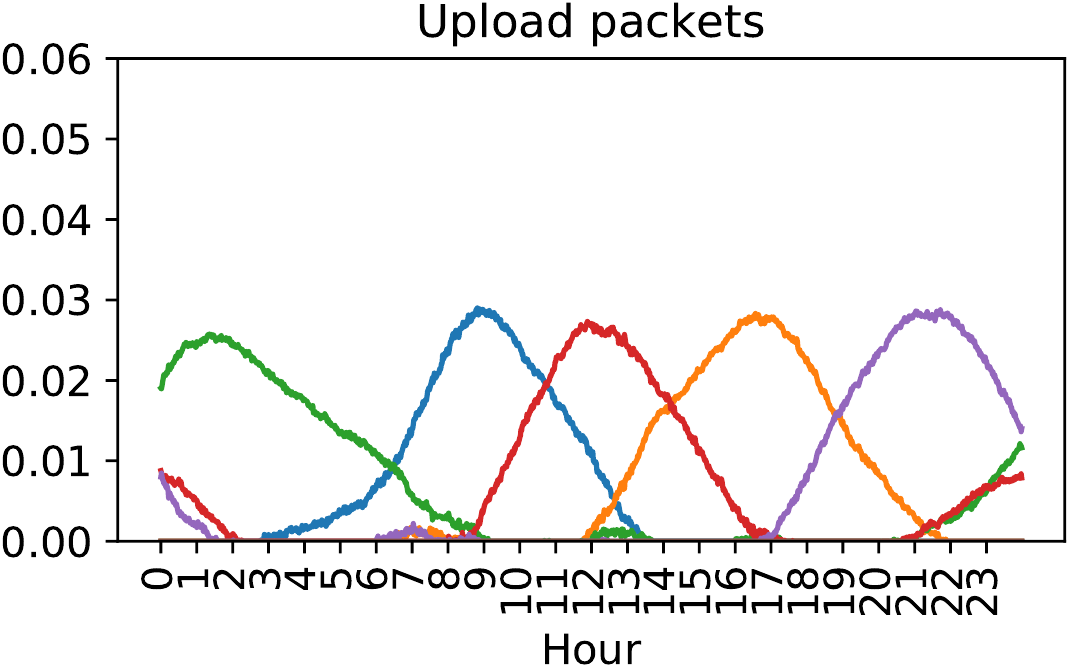}}
\caption{Offline: Factors obtained by PARAFAC.}
\vspace{-4mm}
\label{fig:classification_offline}
\end{figure*}

\section{Application I: DDoS attack detection}\label{sec:classification}

We apply our framework to detect DDoS attacks originated from home devices. 
We consider a dataset with different types of attack vectors (see Table~\ref{tbl:attacks})
obtained by combining home users traffic and attack traffic
measured in laboratory experiments using real malware code. 
Then, we apply a supervised approach to detect when an attack is underway.

\begin{table}[tb]
\centering
\caption{Types of DDoS attacks evaluated}
\label{tbl:attacks}
\renewcommand{\arraystretch}{1.3}
{\scriptsize
\begin{tabular}{|c|c|}
\hline
\textbf{Malware} & \textbf{Attack type (payload size)} \\ \hline
Mirai & UDP flood (1400B) \\ \hline
Mirai & TCP SYN flood (0B) \\ \hline
Mirai & TCP ACK flood (0B) \\ \hline
Mirai & UDP PLAIN flood (1400B) \\ \hline
BASHLITE & UDP flood (1400B) \\ \hline
BASHLITE & TCP SYN flood (0B) \\ \hline
BASHLITE & TCP ACK flood (0B) \\ \hline
\end{tabular}
}
\end{table}

\subsection{Preprocessing}
\label{sec:classification_preprocessing}

We collect upload and download byte and packet rates
per user in a given day (i.e., UD pair), where measurements are performed every minute. 
From these data we obtain the multivariate time series
used as input to the tensor decomposition method.
We consider 18722 time series from 812 users between 19-August-2019 and 22-September-2019.

We obtain our attack dataset using the approach proposed in~\cite{gabriel_ddos}. A brief description of the methodology follows.
First, we randomly choose a fraction of infected homes $q = 0.05$ that participate in
synchronized DDoS attacks, where $q$ is chosen based on the fraction of users
affected by a real attack~\cite{reuters2016telekom}.
Next, we define the attack type (Table~\ref{tbl:attacks}) uniformly at random.
The majority of attacks have a duration of a few minutes~\cite{blenn2017quantifying}.
Therefore, we consider attacks whose duration follows a Gaussian distribution with mean $\mu = 2$ minutes.
We then sample time slots where the synchronized attacks start using a uniform distribution with one attack per day on average.
Finally, we add the attack traffic to the measured traffic of the infected homes.
Preliminarily analysis were made with more attacks per day and with the same fraction of infected homes.
The achieved results present similar detection performance (not shown).

We split our dataset into three different sets. 
The first set (\emph{Tr}$_1$) contains the first week of the dataset
and is used to extract the normal subspace.
We use a second set (\emph{Tr}$_2$) with the following 19 days
to fit an anomaly classifier using residuals extracted from the tensor model.
A third set (\emph{Te}) with the last 9 days of the dataset
is used to evaluate the classifier performance.

In a real-world scenario, it is often difficult to define precisely 
whether a traffic dataset hides embedded network anomalies, especially in the case of malicious anomalies~\cite{lakhina2005mining}.
Therefore, to evaluate the robustness of our method we consider that the whole dataset, 
including the training set \emph{Tr}$_1$ used to obtain the normal subspace before residual extraction, has
some infected users.
The idea is to consider a more realistic scenario, where some hidden anomalies might be present.

Before applying tensor decomposition, we convert our data to logarithmic scale. 
Then, we apply Min-Max normalization on each traffic metric, where
the minimum and maximum values are taken from the training set \emph{Tr}$_1$ and applied to the entire dataset. 
By keeping traffic metrics on a similar scale, we capture the correlations between them and ensure that they have the same impact on the optimization process.

\subsection{Tensor decomposition}

We consider different tensor structures for offline and online scenarios.
In the offline approach our tensor is composed of three modes: (UD $\times$ traffic metric $\times$ minute).
We get model $\mathcal{M} \in \mathbb{R}^{3412 \times 4 \times 1440}$ from the training set \emph{Tr}$_1$
with a total number of UDs equal to 3412.
The application of \emph{Split-Half Validation} validates up to $R = 6$ factors.
(Except otherwise noted, we use $R=6$.)
In the online approach we consider as modes (user $\times$ traffic metrics $\times$ minute),
so the models $\mathcal{M}(t,W) \in \mathbb{R}^{812 \times 4 \times 1440}$ are obtained
with a window size $W=1440$ over the whole dataset, 
that has a total of 812 users.
We consider four metrics: download and upload bytes/packets at every minute by home users. 

We analyze the factors obtained in the offline scenario (model $\mathcal{M}$) to understand the model behavior.
Figures~\ref{fig:traffic_down} and~\ref{fig:traffic_up} 
present the time mode (mode $C$) factors weighted by the loadings 
associated with download and upload byte rate measurements (mode $B$), respectively, 
while Figures~\ref{fig:pkts_down} and~\ref{fig:pkts_up} show the factors weighted by download and upload packet rate loadings.
One factor (represented in gray)
is nearly constant throughout the day.
The remaining factors identify higher network usage at different periods of the day. 
Moreover, the difference in scale between the number of bytes downloaded and uploaded 
is larger than the difference for the number of packets downloaded and uploaded. 
This indicates that connections exchange a similar number of download and upload packets,
but upload packets usually carry less data.

\subsection{Residual extraction}

\begin{figure}[tb]
\centering
\subcaptionbox{\label{fig:residual_up_packets}}{\includegraphics[width=45.9mm]{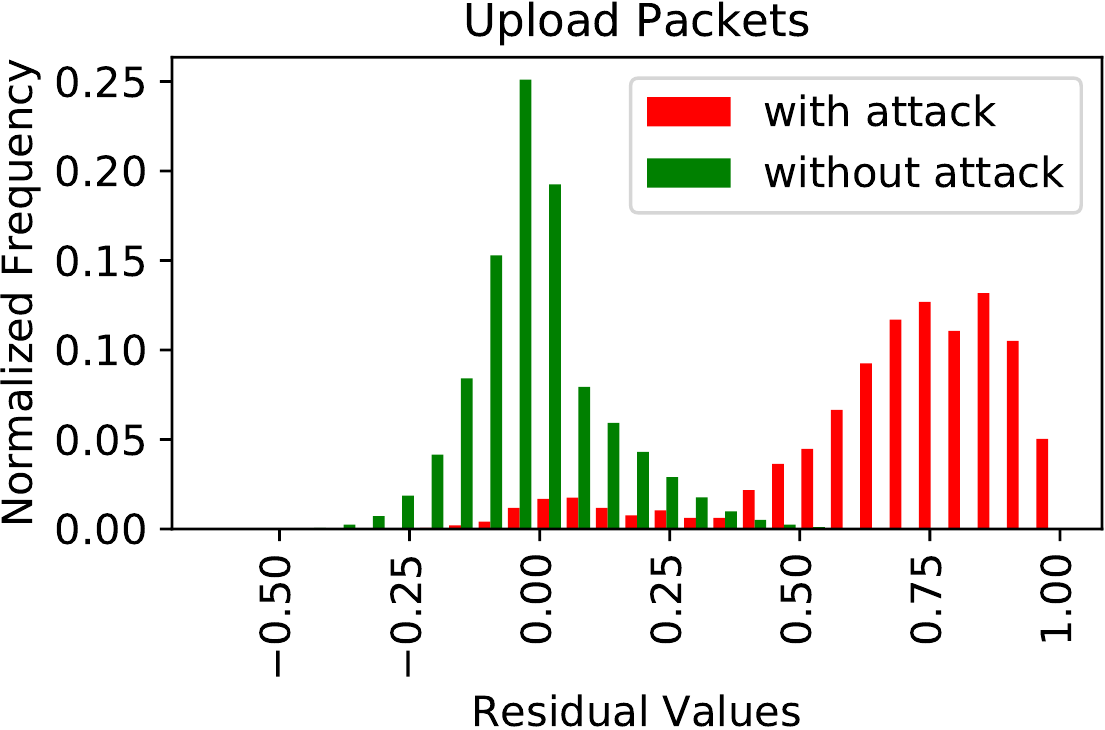}}
\subcaptionbox{\label{fig:residual_diff_packets}}{\includegraphics[width=41.5mm]{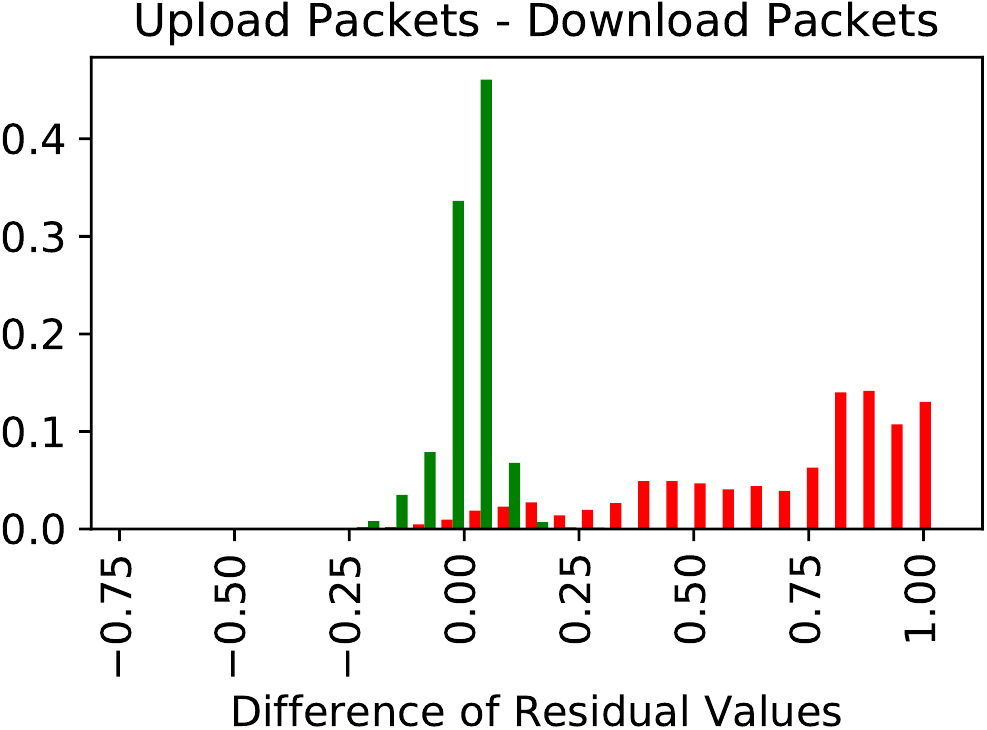}}
\caption{Offline: Histograms of residuals.}
\vspace{-4mm}
\label{fig:residuals}
\end{figure}

We extract residuals for sets \emph{Tr}$_2$ and \emph{Te}
using the residual extraction techniques described in Section~\ref{sec:residual}.
These residuals consist of all traffic metrics for each minute and for each UD/user (offline/online)
and are used as inputs to the classifiers.
The relationship between upload and download traffic can also be an important feature to detect attacks~\cite{gabriel_ddos}.
Therefore, we also consider two more features that express 
the residual difference between upload and download packets and bytes, totaling six features: 
(i) \emph{download bytes}, (ii) \emph{upload bytes}, (iii) \emph{download packets}, (iv) \emph{upload packets}, 
(v) \emph{difference between upload and download bytes} and (vi) \emph{difference between upload and download packets}.
Figures~\ref{fig:residual_up_packets} and~\ref{fig:residual_diff_packets} present the histograms
of the residuals $\tilde{\mathcal{E}}$ retrieved from the offline method 
for the features (iv) \emph{upload packets} and (vi) \emph{difference between upload and download packets}. 
The histograms show that the selected features satisfactorily separates residuals with and without attacks.

\begin{figure}[b]
\centering
\includegraphics[width=65mm]{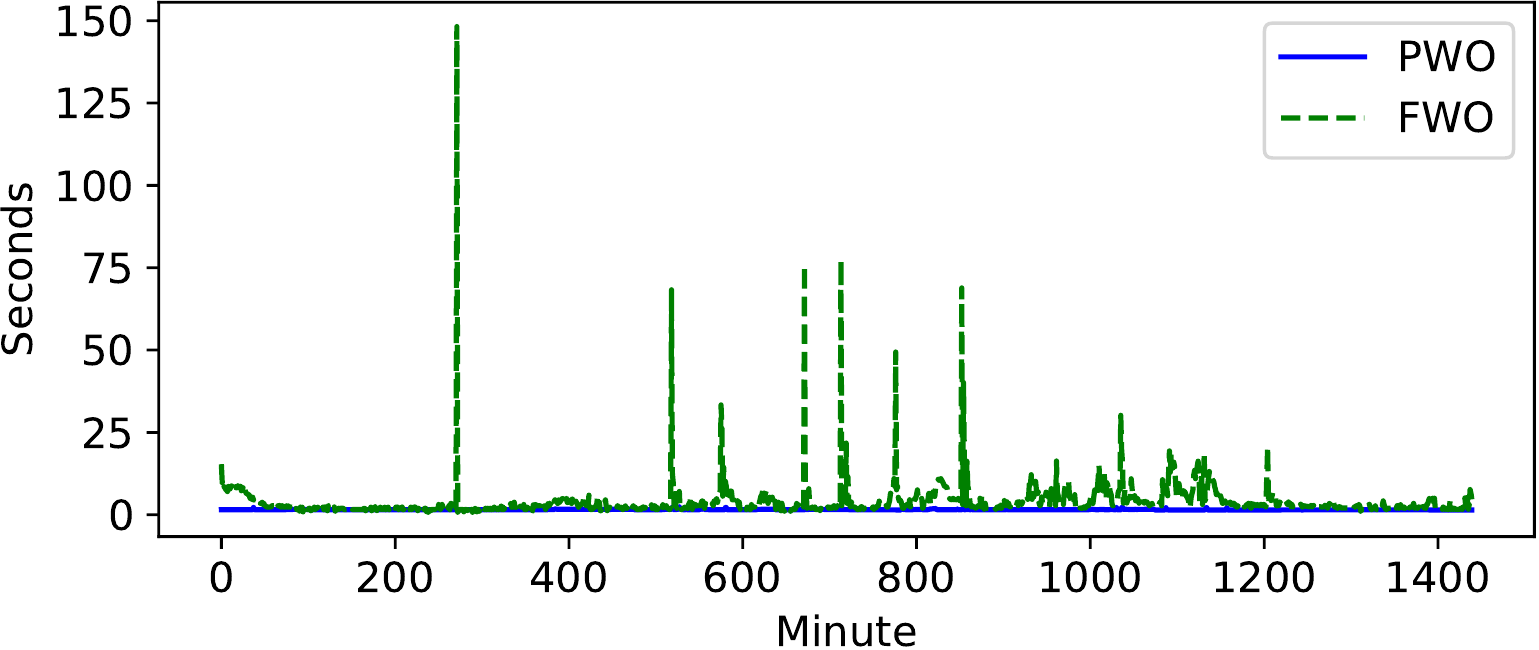}
\vspace{-2mm}
\caption{Run time for FWO and PWO online methods.}
\label{fig:time_compare}
\end{figure}

Figure~\ref{fig:time_compare} shows the run time for both PWO and FWO online methods
for each minute of a day.
Since FWO recomputes all loadings of factor matrix $C(t)$ (time) at every slide of the window (at every minute), 
the time in seconds needed to recompute $\mathcal{M}(t,W)$ varies depending on data
$\mathcal{X}(t,W)$, and can be large enough such as to be inadequate for an online approach. On the other hand, PWO consistenly
requires smaller computational times than FWO.

\subsection{Anomaly classification}

After extracting the residuals we train a classifier to detect when an attack occurs.
To estimate the method's ability to detect attacks we consider five different classifiers,
leveraging features extracted using PARAFAC:
\emph{Logistic Regression}, \emph{Decision Tree}, \emph{Random Forest}, 
\emph{Gaussian Naive Bayes} and \emph{Multi-layer Perceptron}. 
We select the classifier with the best weighted F1 score in a 5-fold cross validation
on the residuals obtained from the training set \emph{Tr}$_2$.
We also consider PCA as an alternative to PARAFAC for comparison purposes. 
As expected,  for the five classifiers considered PARAFAC outperforms PCA as the latter loses 
structural information present in the data after converting tensors into matrices.
The \emph{Random Forest} classifier achieves better results for both methods.
Moreover, preliminary evaluations indicate that models with two factors ($R=2$) perform well. 
Therefore, the results in the sequel are obtained with \emph{Random Forest} and with $R=2$.

Next, we train the \emph{Random Forest} classifier with the residuals obtained from set \emph{Tr}$_2$ 
and we evaluate the results using the residuals of set \emph{Te}. 
Table~\ref{tbl:detection} presents results for two evaluation metrics: Detection Accuracy and Precision.
Detection Accuracy measures the percentage of anomalies detected, 
defined as $\frac{n_d}{n}$, where $n_d$ is the number of detected attacks, 
and $n$ is the total number of attacks in the set \emph{Te}. 
We assume that an attack is detected if an anomaly is identified
in at least one of the time slots that contain the traffic from that attack.
Precision is calculated as follows: $\frac{\textrm{TP}}{\textrm{TP}+\textrm{FP}}$, 
where $\textrm{TP}$ (True Positives) is the time (in minutes) where an attack occurs and is detected,
while $\textrm{FP}$ (False Positives) is defined as the time (in minutes) where an attack is wrongly detected.
Therefore, Precision decreases when the number of False Positives increases.
A model with a better detection rate (higher Detection Accuracy) is critical in scenarios where attacks
have a major impact on the network. 
At the same time, a lower number of false positives (higher Precision) 
decreases the amount of users incorrectly classified as attackers, reducing the chance of a user being wrongly affected
by a countermeasure. For example, a legitimate customer might have its connection blocked if
the classifier falsely reports an attack.

\begin{table}[tb]
\begin{minipage}[t]{.5\linewidth}
\centering
\caption{Random Forest \\ results}
\renewcommand{\arraystretch}{1.3}
{\scriptsize
\begin{tabular}{|l|c|c|}
\hline
\textbf{Model}                          & \textbf{Precision}   &\vtop{\hbox{\strut \textbf{Detection}}\hbox{\strut \textbf{Accuracy}}}\\ \hline
PARAFAC                                 & $0.9891$             & $0.9396$                        \\ \hline
PCA                                     & $0.9709$             & $0.9121$                        \\ \hline
PWO                                     & $0.9718$             & $0.9396$                        \\ \hline
\vtop{\hbox{\strut PWO +}\hbox{\strut Likelihood}}     & $0.9978$             & $0.9725$ \\ \hline
\end{tabular}
}
\label{tbl:detection}
\end{minipage}%
\begin{minipage}[t]{.5\linewidth}
\centering
\caption{Feature importance of  Random Forest}
\renewcommand{\arraystretch}{1.3}
{\scriptsize
\begin{tabular}{|l|c|}
\hline
\textbf{Residual Feature}                        & \textbf{Gini index}     \\ \hline
\vtop{\hbox{\strut Difference up and}\hbox{\strut down packets}} & $0.5207$                \\ \hline
Up packets                              & $0.1489$                \\ \hline
Down packets                            & $0.1348$                \\ \hline
\vtop{\hbox{\strut Difference up and}\hbox{\strut down bytes}}   & $0.1150$                \\ \hline
Down bytes                              & $0.0450$                \\ \hline
Up bytes                                & $0.0356$                \\ \hline
\end{tabular}
}
\label{tbl:feature_importance}
\end{minipage}%
\end{table}

We compare the performance of PARAFAC and PCA models in the offline approach.
Table~\ref{tbl:detection} shows that PARAFAC achieves higher performance for both metrics.
PARAFAC not only detects a higher percentage of attacks but also achieves higher precision, 
with a lower number of false positives.
Moreover, although not reported, our results using PARAFAC show that all types of DDoS attacks 
evaluated have similar Detection Accuracy.

To evaluate the relevance of the six features retrieved from PARAFAC residuals 
we look at the Gini index-based importance metric from the \emph{Random Forest} classifier, 
as shown in Table~\ref{tbl:feature_importance}.
The most important residual feature is the \emph{difference between upload and download packets} (Gini $0.5207$)
followed by \emph{upload packets} (Gini $0.1489$) and \emph{download packets} (Gini $0.1348$). 
We evaluate the classifier using only packet-rate based features and compare the results
against those obtained with PARAFAC for all the features.
The results show that the number of false positives is larger in the first case, 
with Precision decreasing to $0.9780$ using only packet features compared against $0.9891$ using all features.

Table~\ref{tbl:detection} shows that PWO preserves the same Detection Accuracy for online decomposition
in comparison to the offline method, while Precision decays from $0.9891$ (PARAFAC) to $0.9718$ (PWO).
A small decrease in performance is not surprising taking into account that PWO is an online approach 
where the model is constantly updated as soon as a new data stream arrives. 
Nevertheless, PWO still achieves better results for both metrics in comparison to PCA.
We also consider the time to detect an attack in the online scenario.
The detection time of the PWO model is one minute for $86.55\%$ of the detected attacks, 
while $99.41\%$ of the attacks are detected within two minutes.
A short detection time is essential to adopt fast countermeasures and mitigate the impact of an attack.

Figure \ref{fig:residuals} suggests that the histograms of residuals $\tilde{\mathcal{E}}$
follow a two-component mixture of Gaussians (GMM):
one mixture represents residuals when only normal traffic is present and another
when there are attacks. 
(Each GMM has six dimensions, one per feature.
Figure \ref{fig:residuals} shows two of the six dimensions.)
To leverage this observation for classification
purposes, we add two additional features to the online
PWO classifier, namely the likelihoods that residuals are generated by each GMM at
each minute.
(Therefore, eight features are used in total.)
The results reported in the last line of Table~\ref{tbl:detection} indicate that those
two additional features can significantly increase precision and
accuracy.

\subsection{Spatio-temporal correlation}

It is possible to further improve attack detection rates
by correlating the classifier results for each home, since DDoS
attacks are synchronized by nature.
Mendonça et al.~\cite{gabriel_ddos} propose a Bayesian decision problem using MAP criterion
to detect synchronized attacks with high probability.
Using the model parameterized with our results for the PWO online method ($|\mathcal{H}| = 812, P_{D} \approx 0.0014, p_{fp} \approx 2.64 \cdot 10^{-6}, p_{rc} \approx 0.8266, q = 0.05$)
yields $m_0 \approx 4.21$. Therefore, the model considers that a synchronized attack is happening if at least 5
users report an attack. The spatio-temporal correlation presents a great performance with the probability of false alarms 
(Type I error) equals $3.73 \cdot 10^{-16}$ and the probability of missing a synchronized attack (Type II error) equals $9.11 \cdot 10^{-11}$.

%% file: tex/clustering.tex
\section{Application II: Detecting network degradation intervals}\label{sec:clustering}

We apply our methodology to the detection of degradation intervals in the ISP network.
In the absence of reliable labels to identify anomalies and evaluate the results quantitatively,
we rely on unsupervised clustering over residuals extracted by the offline method to group events with
similar behavior.
An application example of the method is to automatically identify potential network problems
affecting multiple users and to show the regions with poor performance.

\subsection{Preprocessing}

We use both latency and loss time series measured at one-minute intervals as the main performance metrics of interest. These 
metrics were collected on 2964 home routers between 19-August-2019 and 22-September-2019.
Both metrics are obtained by sending a train of 100 ICMP packets at 10 millisecond intervals to a server located in the ISP network. 
Latency and loss measurements can be affected by home network user traffic~\cite{sam-results}.
Therefore, we do not consider for the analysis the value of minute samples whose cross-traffic is greater than a threshold $\theta$.
Based on the users with the lowest nominal capacity in our dataset we set this threshold as $\theta = 2.5$~Mbps.
After filtering out cross-traffic, we only consider time series with at least  $\eta = 1000$ samples.

For some network failures, e.g. link failures, the communication between a client and the measurement server can be 
disrupted and no measurement samples are recorded. Therefore, it is possible to infer periods of network unavailability from the lack
of measurement samples, specially when multiple users do not report measurement results simultaneously.
Hence, we encode every minute bins for which loss measurements are missing as having packet loss rate equal to 1.
For the analysis, we remove samples with cross-traffic above the threshold $\theta$ and periods when the measurement
server is offline.
This introduces missing samples that PARAFAC can easily handle.
Finally, we use the log of each sample as input for the tensor decomposition.
We considered 50282 multivariate time series with latency and loss values of one-minute granularity.

\subsection{Tensor decomposition and residual extraction}

We model the measurement data as a tensor with three modes: (UD $\times$ network metric $\times$ minute). 
Since we use daily time series of latency/loss measurements as inputs, we get a third-order tensor $\mathcal{X} \in \mathbb{R}^{50282 \times 2 \times 1440}$. 
We use the \emph{Split-Half Validation} method to set the number of factors $R = 4$.

We use metrics obtained from three different residual time series: latency residuals, 
loss residuals in which  
samples with packet loss fraction equal to 1 are removed and loss residuals that includes
all samples. Each metric can be used to detect a different type of
anomaly, such as network congestion and link or equipment failures. We extract three statistics for each residual time series: mean, standard deviation and $95^{th}$ percentile, totaling nine features.

\subsection{Anomaly clustering}

\subsubsection{Clustering results}
We use $K$-Means for clustering the residuals due to its simplicity and interpretability.
To select the number of clusters, we use the Elbow Method~\cite{lakhina2005mining}. Five clusters are chosen.
Before clustering the data we apply the $z$-score normalization to avoid that any feature dominates the analysis due to scaling.

To investigate the meaning of each cluster we summarize all
time series assigned to each cluster considering three metrics:
latency, loss and amount of missing samples. To evaluate
the latency per cluster we apply a normalization obtained by
subtracting out the lowest value from each daily time series
in order to infer packet queuing times during congestion
periods.

\begin{figure}[t]
\centering
\subcaptionbox{Latency  \label{fig:latency_per_cluster}}{\includegraphics[width=43.5mm]{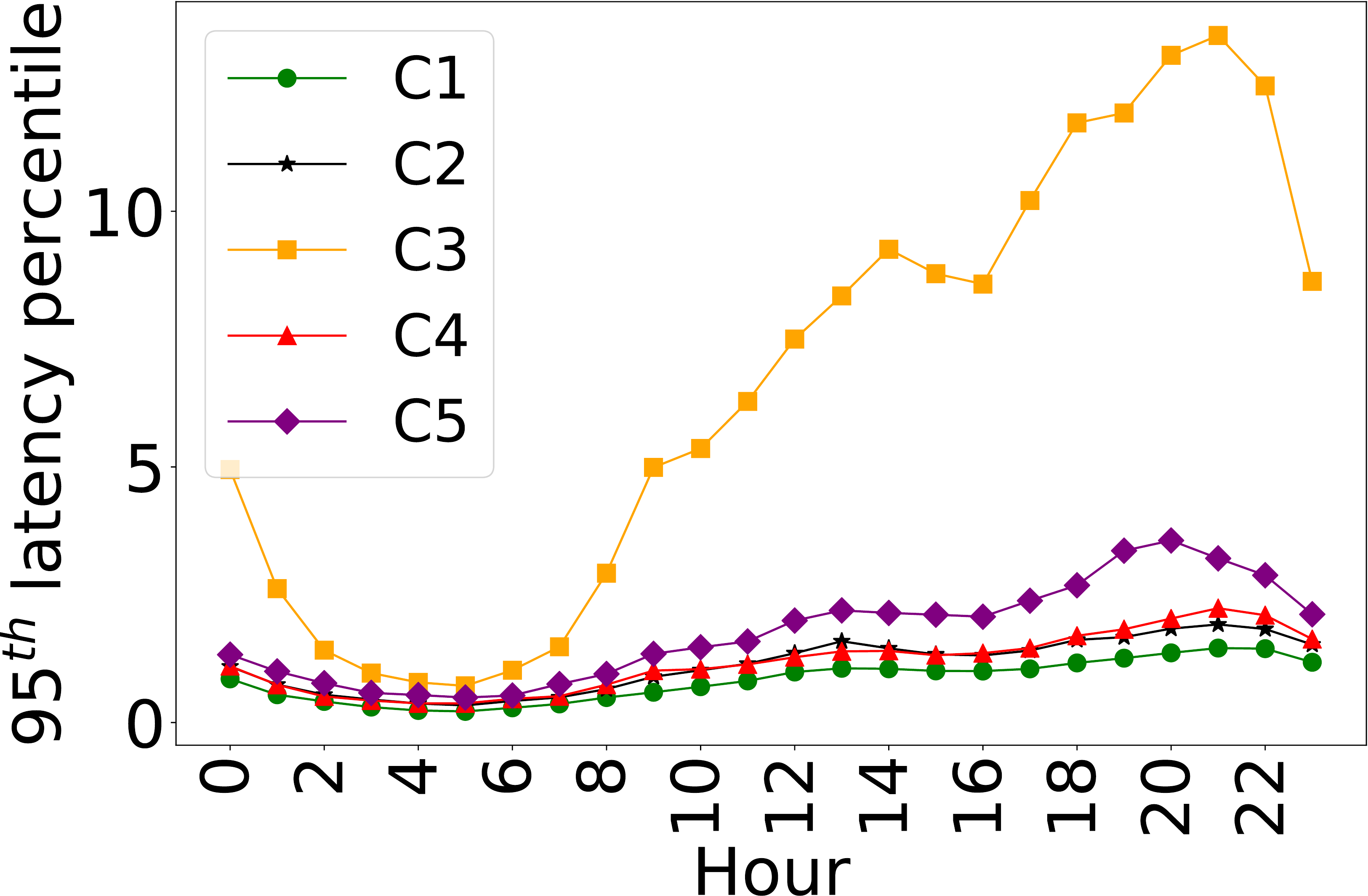}}
\subcaptionbox{Fraction of missing samples  \label{fig:unavailability_per_cluster}}{\includegraphics[width=43.5mm]{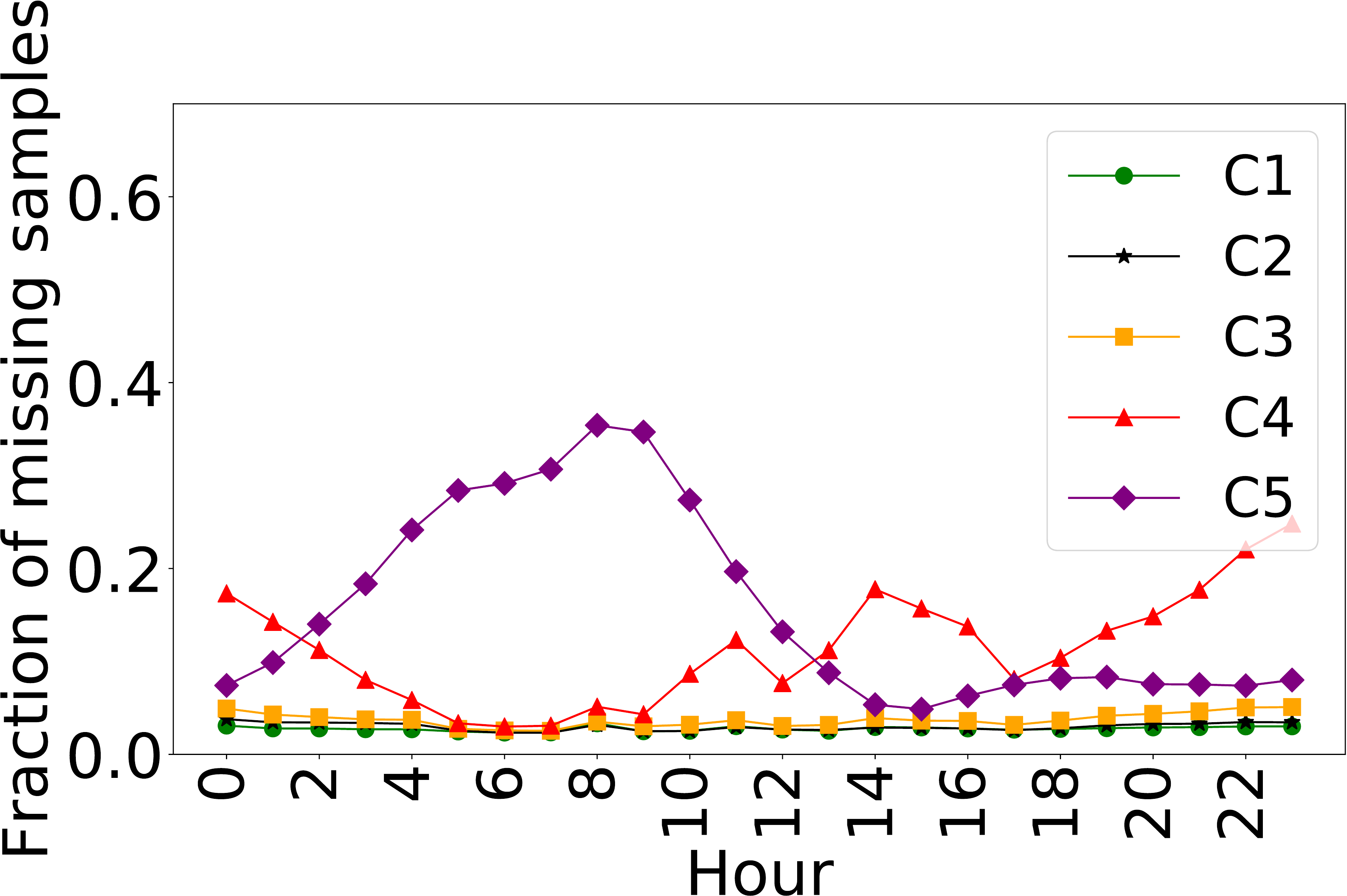}}
\par\bigskip
\subcaptionbox{Loss\label{fig:loss_per_cluster}}{\includegraphics[width=43.5mm]{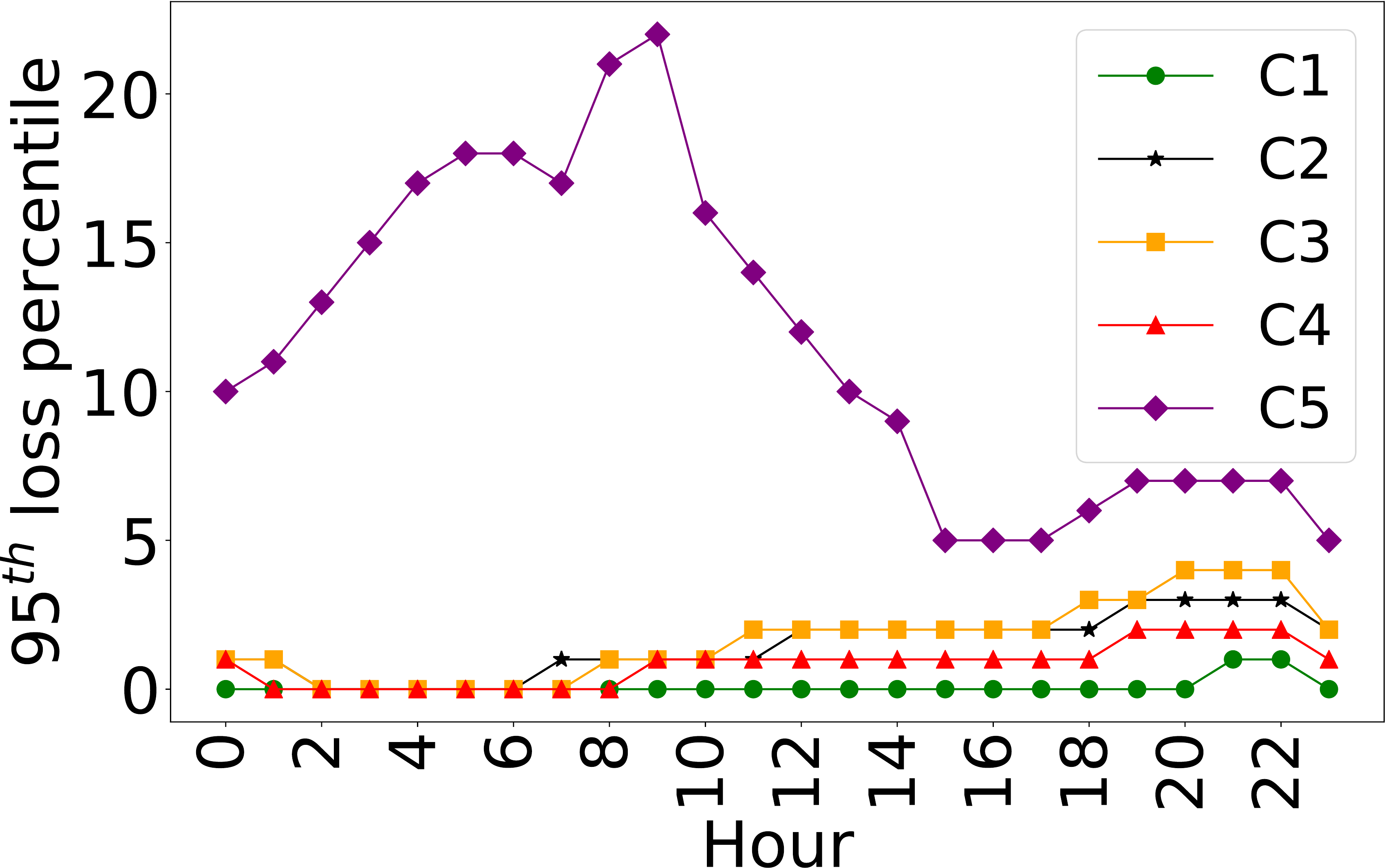}}
\caption{Summarized time series for each cluster}
\vspace{-4mm}
\label{fig:clustering_coords}
\end{figure}

Figures~\ref{fig:latency_per_cluster}, \ref{fig:unavailability_per_cluster} and \ref{fig:loss_per_cluster} show the summaries obtained for each metric in all clusters.
In summary, users of cluster C1 experience  good quality of service (low latency, low loss, low unavailability). Cluster C2 contains time series with moderate losses but low unavailability and low latency. Cluster C3 contains time series with high latency and moderate losses. The time series of cluster C4 are from users that experience high unavailability periods, while cluster C5 contains time series with both high unavailability and loss rates.
Note that we cluster UD pairs, i.e., daily measurement time series of different users. Therefore, a user can be assigned to different clusters at different days, as we show in the next section.

\subsubsection{Spatial correlation}
To identify periods experiencing performance degradation affecting multiple geographically ``close'' users,
we spatially correlate the clustering results and ISP topology information.
The spatio-temporal correlation algorithm assumes that the routes between home-routers and the measurement server are static during 
each measurement interval. Consequently, the network topology can be represented by a tree structure at each measurement interval.
We expect clients that share the same ISP network paths should exhibit similar performance inside the ISP network in terms of 
congestion and failures. We analyze the fraction of users assigned to each cluster at each day of the dataset.

\begin{figure}[t]
    \centering
    \includegraphics[width=43.5mm]{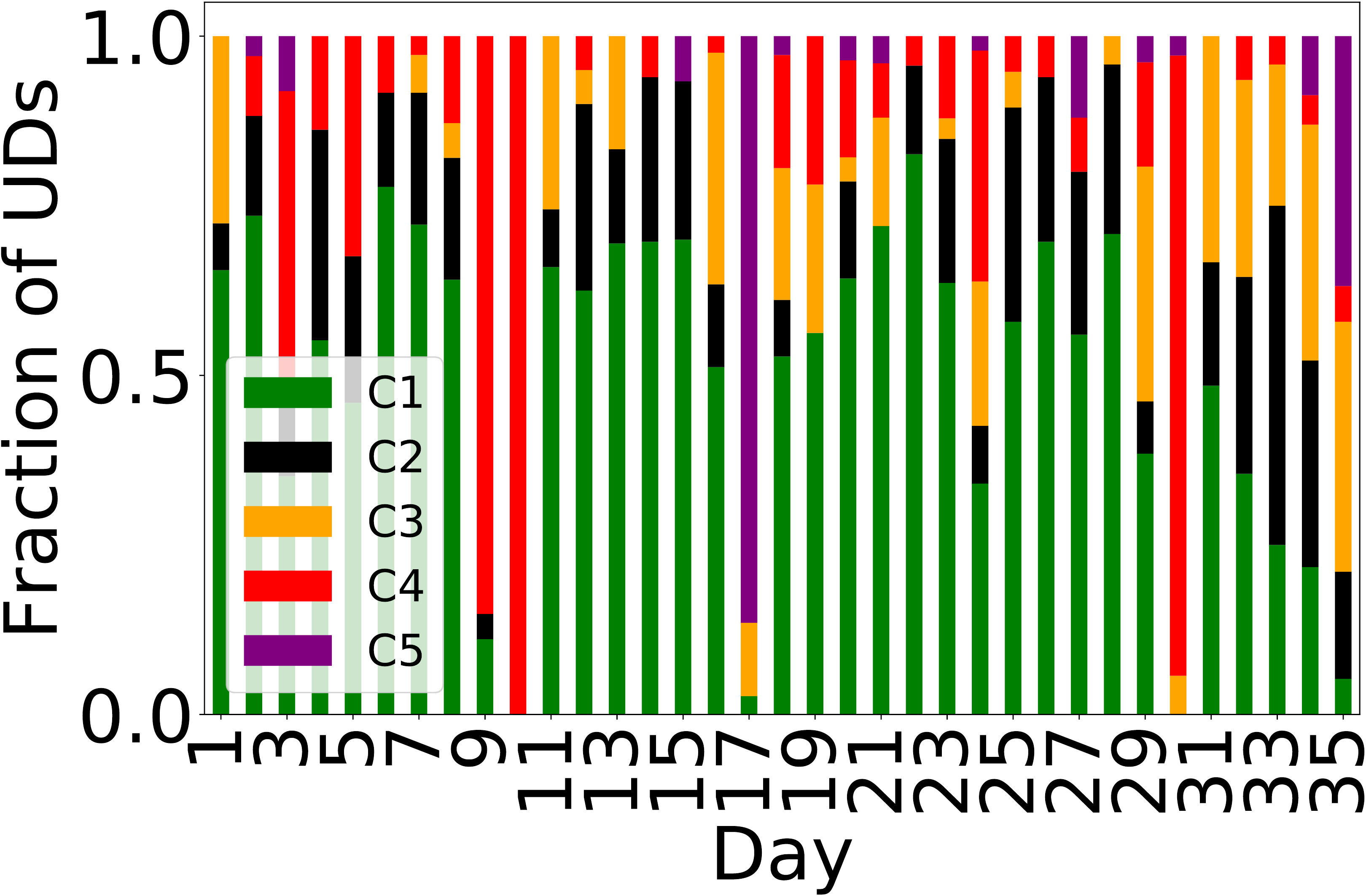}
    \vspace{-2mm}
    \caption{Spatial correlation example}
\label{fig:topology_cluster_daily}
\vspace{-4mm}
\end{figure}

We exemplify the results of the spatial correlation for a specific region of the network.
Similar results are obtained for other network regions. 
Figure~\ref{fig:topology_cluster_daily} shows the daily fraction of UD pairs per cluster.
Usually the majority of users are associated with cluster C1 and few losses are observed.
However a large number of users are associated with cluster C4 at day 10,
when multiple time series have missing samples between 1 P.M. and 5 P.M. 
Another type of event detected by the spatial correlation occurs on day 17,
when periods with missing samples between 6 A.M. and 8 A.M. and high losses between 7 P.M. and 9 P.M.
were observed and several users are assigned to cluster C5.

\begin{figure}[ht]
\centering
\subcaptionbox{Region A\label{fig:cluster_summary_good_region}}{\includegraphics[width=43.5mm]{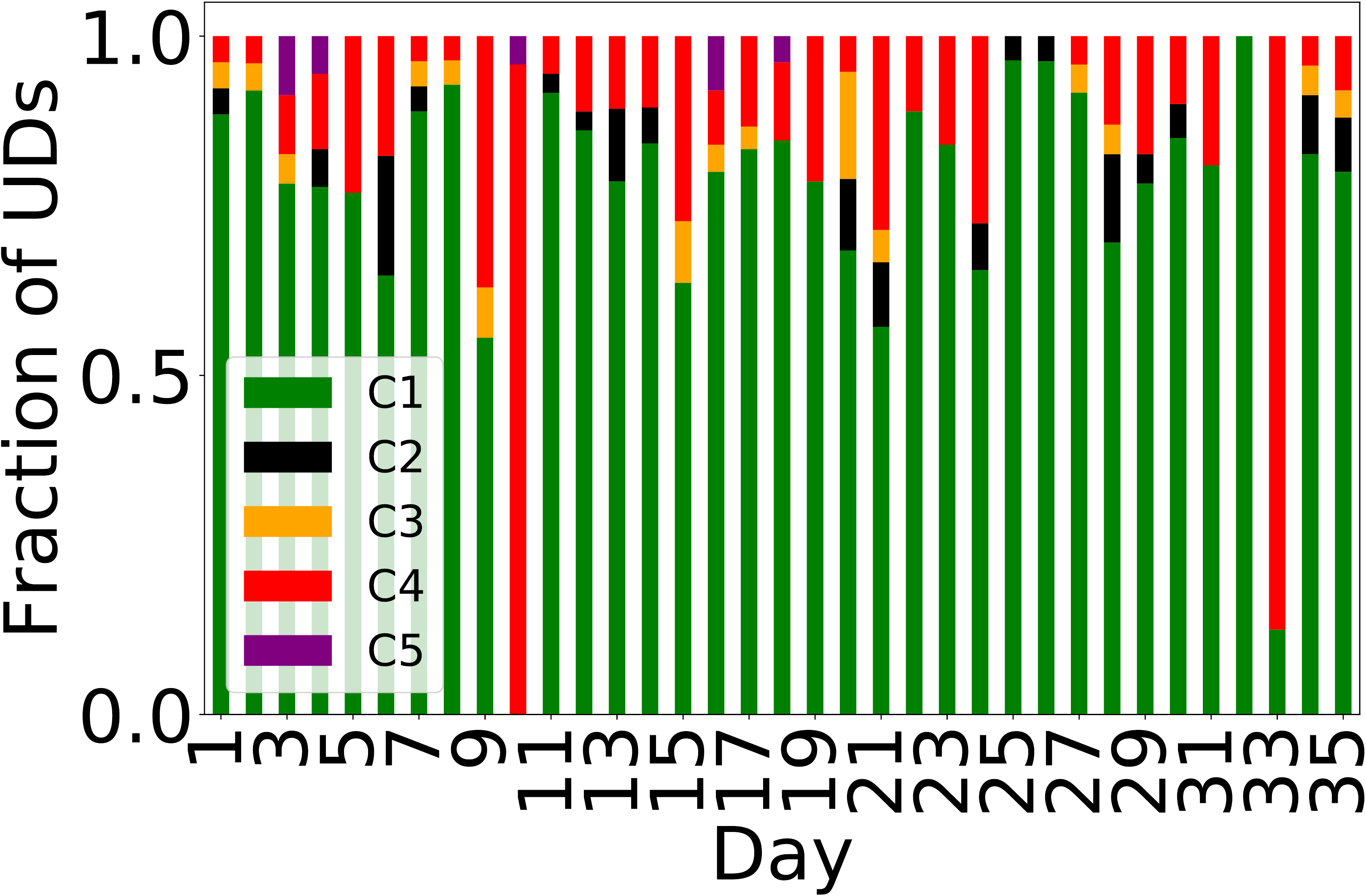}}
\subcaptionbox{Region B\label{fig:cluster_summary_bad_region}}{\includegraphics[width=43.5mm]{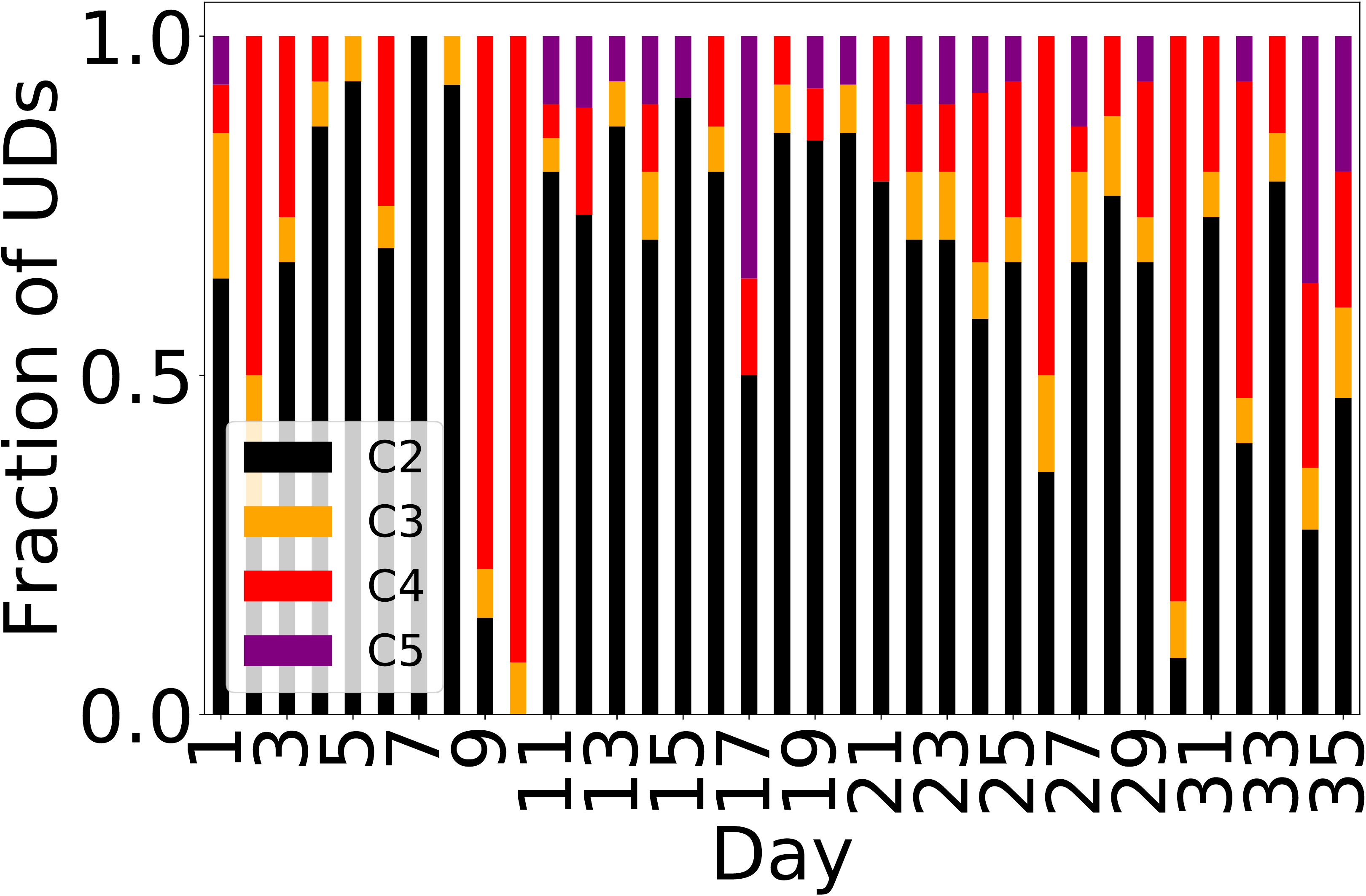}}
\caption{Summary of network performance obtained from residual clustering}
\vspace{-4mm}
\label{fig:cluster_summary}
\end{figure}

The framework results can summarize the quality of each network region based on the number of UD pairs assigned to the cluster representing good performance (cluster C1). 
Figure~\ref{fig:cluster_summary} presents the clustering results in two different portions
of the network. 
It can be seen that one region consistently presents a high fraction of users associated with better performance 
(Figure~\ref{fig:cluster_summary_good_region}), 
although performance degradation periods can be identified
on days 10 and 33. 
At the same time, Figure~\ref{fig:cluster_summary_bad_region} shows a region where no users are assigned to cluster C1. 

%% file: tex/conclusion.tex
\section{Conclusion} \label{sec:concl}

In this work, we propose a method based on tensor decomposition to detect network anomalies. 
We apply the PARAFAC method and extract the residuals obtained by the model in order to detect abnormal behavior.
We also propose a new online tensor decomposition method that efficiently extracts the normal subspace and detects
anomalies with good performance.
We show the flexibility of our method, using two different applications as examples.
First, we consider DDoS attack detection using supervised techniques.
The results show that we can obtain high values for Detection Accuracy and Precision using different classifiers.
Besides, our method has better performance and robustness when compared to PCA.
Then, we use the proposed methodology to identify periods of network performance degradation through an unsupervised approach.
The method is able to identify periods of degradation affecting several customers and
identify QoS related problems on different parts of an ISP's topology.